\documentclass[aps,prl,superscriptaddress,twocolumn,showpacs,amsmath,amssymb,floatfix]{revtex4}

\usepackage{graphicx}
\usepackage{dcolumn}
\usepackage{bm}

\newcommand{\keemm}{K_{L}\to e^+e^-\mu^+\mu^-\,}
\newcommand{\eemm}{ee\mu\mu\,}
\def\ee{e^+e^-}
\def\mm{\mu^+\mu^-}
\newcommand{\keeee}{K_{L}\to e^+e^-e^{+}e^{-} \,}
\newcommand{\kmm}{K_{L}\to \mu^+\mu^-\,}
\newcommand{\kwrongsign}{K_{L}\to e^{\pm}e^{\pm}\mu^{\mp}\mu^{\mp}\,}
\newcommand{\kmmg}{K_{L}\to \mu^+\mu^-\gamma \,}
\newcommand{\kpmzdal}{K_{L}\to \pi^+\pi^-\pi^{0}_{D} \,}
\newcommand{\pt}{P_{t}^{2}}
\newcommand{\klgg}{K_{L}\,\gamma^\ast\,\gamma^\ast \,}
\newcommand{\aks}{\alpha_{K^{\ast}}}

\newcommand{\UAz}{University of Arizona, Tucson, Arizona 85721}
\newcommand{\UCLA}{University of California at Los Angeles, Los Angeles,
                    California 90095} 
\newcommand{\UCSD}{University of California at San Diego, La Jolla,
                   California 92093} 
\newcommand{\Campinas}{Universidade Estadual de Campinas, Campinas, 
                       Brazil 13083-970}
\newcommand{\EFI}{The Enrico Fermi Institute, The University of Chicago, 
                  Chicago, Illinois 60637}
\newcommand{\UB}{University of Colorado, Boulder, Colorado 80309}
\newcommand{\ELM}{Elmhurst College, Elmhurst, Illinois 60126}
\newcommand{\FNAL}{Fermi National Accelerator Laboratory, 
                   Batavia, Illinois 60510}
\newcommand{\Osaka}{Osaka University, Toyonaka, Osaka 560-0043 Japan} 
\newcommand{\Rice}{Rice University, Houston, Texas 77005}
\newcommand{\SaoPaolo}{Universidade de Sao Paolo, Sao Paolo, Brazil 05315-970}
\newcommand{\UVa}{The Department of Physics and Institute of Nuclear and 
                  Particle Physics, University of Virginia, 
                  Charlottesville, Virginia 22901}
\newcommand{\UW}{University of Wisconsin, Madison, Wisconsin 53706}

\begin{document}


\title{Measurements of the Decay $\keemm$}

\affiliation{\UAz}
\affiliation{\UCLA}
\affiliation{\UCSD}
\affiliation{\Campinas}
\affiliation{\EFI}
\affiliation{\UB}
\affiliation{\ELM}
\affiliation{\FNAL}
\affiliation{\Osaka}
\affiliation{\Rice}
\affiliation{\SaoPaolo}
\affiliation{\UVa}
\affiliation{\UW}

\author{A.~Alavi-Harati}  \affiliation{\UW}
\author{T.~Alexopoulos}   \affiliation{\UW}
\author{M.~Arenton}       \affiliation{\UVa}
\author{K.~Arisaka}       \affiliation{\UCLA}
\author{R.F.~Barbosa}     \affiliation{\SaoPaolo}
\author{A.R.~Barker}      \affiliation{\UB}
\author{M.~Barrio}        \affiliation{\EFI}
\author{L.~Bellantoni}    \affiliation{\FNAL}
\author{A.~Bellavance}    \affiliation{\Rice}
\author{E.~Blucher}       \affiliation{\EFI}
\author{G.J.~Bock}        \affiliation{\FNAL}
\author{C.~Bown}          \affiliation{\EFI}
\author{S.~Bright}        \affiliation{\EFI}
\author{E.~Cheu}          \affiliation{\UAz}
\author{R.~Coleman}       \affiliation{\FNAL}
\author{M.D.~Corcoran}    \affiliation{\Rice}
\author{B.~Cox}           \affiliation{\UVa}
\author{A.R.~Erwin}       \affiliation{\UW}
\author{C.O.~Escobar}     \affiliation{\Campinas}
\author{R.~Ford}          \affiliation{\FNAL}
\author{A.~Glazov}        \affiliation{\EFI}
\author{A.~Golossanov}    \affiliation{\UVa}
\author{P. Gouffon}       \affiliation{\SaoPaolo}
\author{J.~Graham}        \affiliation{\EFI}
\author{J.~Hamm}          
\altaffiliation[To whom correspondence should be addressed.]{ }
\affiliation{\UAz}
\author{K.~Hanagaki}      \affiliation{\Osaka}
\author{Y.B.~Hsiung}      \affiliation{\FNAL}
\author{H.~Huang}         \affiliation{\UB}
\author{V.~Jejer}         \affiliation{\UVa}
\author{D.A.~Jensen}      \affiliation{\FNAL}
\author{R.~Kessler}       \affiliation{\EFI}
\author{H.G.E.~Kobrak}    \affiliation{\UCSD}
\author{K.~Kotera}        \affiliation{\Osaka}
\author{J.~LaDue}         \affiliation{\UB}
\author{N.~Lai}           \affiliation{\EFI}
\author{A.~Ledovskoy}     \affiliation{\UVa}
\author{P.L.~McBride}     \affiliation{\FNAL}

\author{E.~Monnier}
   \altaffiliation[Permanent address ]{C.P.P. Marseille/C.N.R.S., France}
   \affiliation{\EFI}

\author{K.S.~Nelson}     \affiliation{\UVa}
\author{H.~Nguyen}       \affiliation{\FNAL}
\author{V.~Prasad}       \affiliation{\EFI}
\author{X.R.~Qi}         \affiliation{\FNAL}
\author{B.~Quinn}        \affiliation{\EFI}
\author{E.J.~Ramberg}    \affiliation{\FNAL}
\author{R.E.~Ray}        \affiliation{\FNAL}
\author{E. Santos}       \affiliation{\SaoPaolo}
\author{K.~Senyo}        \affiliation{\Osaka}
\author{P.~Shanahan}     \affiliation{\FNAL}
\author{J.~Shields}      \affiliation{\UVa}
\author{W.~Slater}       \affiliation{\UCLA}
\author{N.~Solomey}      \affiliation{\EFI}
\author{E.C.~Swallow}    \affiliation{\EFI}\affiliation{\ELM}
\author{S.A.~Taegar}     \affiliation{\UAz}
\author{R.J.~Tesarek}    \affiliation{\FNAL}
\author{P.A.~Toale}      \affiliation{\UB}
\author{A.~Tripathi}     \affiliation{\UCLA}
\author{R.~Tschirhart}   \affiliation{\FNAL}
\author{Y.W.~Wah}        \affiliation{\EFI}
\author{J.~Wang}         \affiliation{\UAz}
\author{H.B.~White}      \affiliation{\FNAL}
\author{J.~Whitmore}     \affiliation{\FNAL}
\author{M.~Wilking}      \affiliation{\UB}
\author{B.~Winstein}     \affiliation{\EFI}
\author{R.~Winston}      \affiliation{\EFI}
\author{E.T.~Worcester}  \affiliation{\UCLA}\affiliation{\EFI}
\author{T.~Yamanaka}     \affiliation{\Osaka}
\author{R.F.~Zukanovich} \affiliation{\SaoPaolo}

\begin{abstract}
The KTeV experiment at Fermilab has isolated a total of 132 events from the 
rare decay $\keemm$, with an estimated background of 0.8 events.  The 
branching ratio of this mode is determined to be 
$(2.69\pm 0.24_{stat}\pm 0.12_{syst})\times 10^{-9}$, 
with a radiative cutoff of $M_{\eemm}^2/M_{K}^2 > 0.95$.  
The first measurement using this mode of the parameter $\alpha$ from the 
D'Ambrosio, Isidori, and Portol\`{e}s model of the $\klgg$ vertex yields 
a result of $-1.59\pm 0.37$, consistent with values obtained from other 
decay modes.  Because of the limited statistics,
no sensitivity is found to the DIP parameter $\beta$.  
The magnitude of the angular distribution asymmetry between the
$\ee$ and $\mm$ planes,
indicative of a $CP-$violating contribution to the decay, 
is found to be consistent with zero.  We set a 90\% C.L. upper limit of 
$4.12\times 10^{-11}$ on the branching ratio of the lepton flavor--violating 
mode $\kwrongsign$, a factor of three improvement over the current limit 
from the KTeV experiment.
\end{abstract}

\maketitle

The rare decay $\keemm$ offers the most direct
means for studying the 
dynamics of the $\klgg$ vertex.  This information is useful for 
models that relate the $\kmm$ branching ratio to $\rho$, the real part 
of the CKM matrix element 
$V_{td}$~\cite{Wolfenstein:1983yz, Belanger:1991ur, Buras:1998fb}.  
This decay mode can also be used to determine the presence of 
any $CP-$violating 
contributions to the $\klgg$ interaction~\cite{Uy:1991hu}.  Additionally, 
a search for the lepton flavor--violating counterpart $\kwrongsign$ provides a 
constraint on physics beyond the Standard Model.


In the model of 
D'Ambrosio, Isidori, and Portol\`{e}s (DIP)~\cite{D'Ambrosio:1998jp},  the 
$\klgg$ form factor can be written as
\begin{eqnarray}
\label{eq:dipff}
\nonumber f\left(q_1^2,q_2^2\right) = 
1&+&\alpha\left(\frac{q_1^2}{q_1^2-M_\rho^2}+\frac{q_2^2}{q_2^2-M_\rho^2}\right)\\
&+&\beta\frac{q_1^2\,q_2^2}{\left(q_1^2-M_\rho^2\right)\left(q_2^2-M_\rho^2\right)}.
\end{eqnarray}
Here, $q_1$ and $q_2$ are the momenta of the two virtual photons,
and $M_\rho$ is the mass of the $\rho$ vector meson. 
In this model, $\alpha$ and $\beta$ are two arbitrary real parameters and
are expected to be of order one. The determination
of both $\alpha$ and $\beta$ is possible through the decay $\keemm$
by examining the dilepton invariant masses and the
integrated decay rate. 
Knowledge of the $\klgg$ form factor is important for understanding the 
long distance contributions to $\kmm$ and extracting the value of 
$\rho$~\cite{D'Ambrosio:1998jp}.

Two measurements have been made of the linear DIP parameter $\alpha$ to date, 
both by the KTeV collaboration.  From the mode $\kmmg$, the shape of the 
dimuon invariant mass distribution ($M_{\mu\mu}$) and the measured branching 
ratio have been used to determine $\alpha = -1.54\pm 
0.10$~\cite{Alavi-Harati:2001wd}.  A fit to the dielectron mass 
distribution ($M_{ee}$) from $\keeee$ determines 
$\alpha=-1.1\pm 0.6$~\cite{Alavi-Harati:2001ab}, where the larger
error results from the smaller $q^2$ of the dielectron
distribution.  No measurements have yet been made of the quadratic DIP 
parameter $\beta$. Because the effects of the $\beta$ parameter are most 
significant in the region where both $q_1^2$ and $q_2^2$ are large, the decay
$\keemm$ represents the best means for determining $\beta$. 

KTeV, a fixed target experiment located at Fermilab, collected rare decay 
data during run periods in 1997 and 1999.  Forty three $\keemm$ events were 
observed in the 1997 data, leading to a published branching ratio of 
$(2.62\pm 0.40_{stat}\pm 0.17_{syst})\times 
10^{-9}$~\cite{Alavi-Harati:2001tk}.  However, no attempt has been made 
to extract form factor information from this limited dataset.  The results 
presented in this Letter are based on a reanalysis of the 1997 KTeV dataset, 
combined with the analysis of new data collected during the 1999 run.

The two parallel $K_L$ beams used by KTeV are created by focusing 800 GeV/c 
protons from the Tevatron onto a BeO target.  A 65 m long vacuum region, 
starting 94 m downstream from the target, defines the fiducial region for 
kaon decays.  Charged particles are detected with a spectrometer system 
consisting of four drift chambers and an analysis magnet.  The hit position 
resolution in the chambers is approximately 100 $\mu m$, while the overall 
momentum resolution is just over 1\% in the range of interest.  The 
transverse momentum kick from the magnet was lowered by 25\% for the 1999 
run for the purpose of increasing acceptance for four--track decay modes.  

Downstream of the spectrometer system are two trigger hodoscope planes, 
followed by an electromagnetic (EM) calorimeter.  This $1.9\times 1.9$ m$^2$ 
array of 3100 pure CsI crystals has a resolution of under 1\% in the energy 
range of interest.  Photon vetos are located along the decay region
to reject decays from particles that would miss the CsI calorimeter.

Behind the calorimeter are a 10 cm thick lead wall and 4 m of steel, 
the last 3 m of which serve as the neutral beam dump.  A muon hodoscope (MU2) 
consisting of 56 overlapping scintillator paddles is located behind the dump.  
Following MU2 is another 1 m thick steel filter, behind which are two 
scintillator planes, one oriented horizontally and the other vertically.  
Known as MU3Y and MU3X, these planes are used for muon identification and
have 15 cm segmentation.  The momentum threshold for muons to reach the MU3 
bank has been measured to be 7 GeV/c.  The lead and steel filters add up 
to a total of 31 hadronic interaction lengths.  A more detailed description 
of the KTeV detector can be found 
elsewhere~\cite{Alavi-Harati:1999zr, Alavi-Harati:1999hd}.

The trigger requires hits in the upstream drift chambers and the 
trigger hodoscope planes consistent with at least two charged tracks.  
During the 1997 run, at least two hits were required in both MU3X and MU3Y.  
This condition was loosened during the 1999 run to allow for one missing hit.  
This change accepts events in which the muons are well separated in one view 
but happen to strike the same paddle in the other view.  To counter the 
increased trigger rate from this change, the minimum number of calorimeter 
clusters with at least 1 GeV of energy was raised from one in 1997 to two 
in 1999~\cite{Bown:1996qv}.  If at least 0.5 GeV of energy is found in any 
of the photon vetos, the event is discarded.  Events in which at least three 
tracks form a loosely defined vertex are tagged as candidate signal events.

During the analysis stage, 
the ratio $E/P$ is used for particle identification, 
where $E$ is the energy 
deposited by the track in the EM calorimeter, and $P$ is the track momentum as 
measured by the spectrometer.  Tracks with $0.95<E/P<1.05$ 
are identified as electrons.  Tracks are 
identified as muons if they have $E/P<0.8$, deposit less than 1.5 GeV in 
the calorimeter, 
have momentum greater than 7 GeV/c, 
and hit at least 2 out of the 3 muon identification planes 
(MU2, MU3X, and MU3Y).  
Events are accepted if they contain exactly four tracks, with oppositely 
charged electron and muon pairs.

Additional requirements imposed on the dataset include cuts on the 
reconstructed kaon momentum (20 GeV/c $<P_K<$ 220 GeV/c) and the $z$ position 
of the reconstructed vertex (90 m $<z_{vtx}<$ 158 m).  Because the detector 
acceptance for $K_L$ decays falls off quickly outside of these ranges, any 
events observed outside of the boundaries are most likely misreconstructed 
$K_L$, $K_S$, or hyperon decays.  
To further reduce the number of misreconstructed 
and background events, a cut is made at $\pt < 250$ MeV$^2$/c$^2$, where 
$P_t$ is the transverse component of the reconstructed kaon momentum relative 
to the kaon line of flight.  
The signal mass region is defined to be 482 MeV/c$^2 <M_{ee\mu\mu}<$ 
512 MeV/c$^2$.

Three sources of background are considered.  The decay $\kpmzdal$ (where 
$\pi^0_D$ signifies the Dalitz decay $\pi^0\to e^+e^-\gamma$) could appear as 
signal if the charged pions decay to muons in flight, or punch through the 
muon filter and fire MU3.  These events are removed from the dataset by 
cutting on an extra EM cluster in the calorimeter, indicative of the photon 
from the Dalitz decay of the $\pi^0$.  Two simultaneous 
$K_L\to \pi^\pm \mu^\mp \nu$ ($K_{\mu 3}$) decays could also simulate signal 
if both pions hadronically interact and shower in the calorimeter, mimicking 
electrons.  As two separate two--track decays are unlikely to form a good 
four--track vertex, a cut on four--track vertex quality is successful in 
removing this background.

\begin{figure}
\scalebox{0.475}{\includegraphics{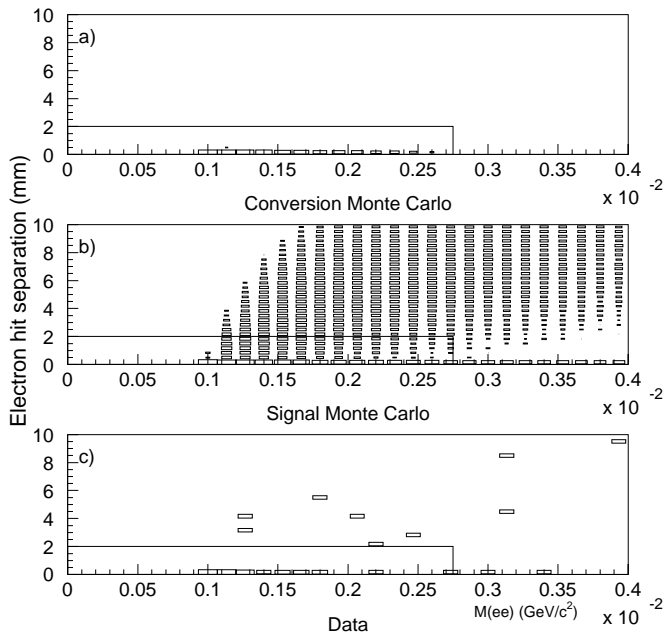}}
\caption{\label{fig:conversions} Electron hit separation at the first drift 
chamber (in millimeters) vs. $M_{ee}$ (in GeV/c$^2$) for a) $\kmmg$ 
conversion Monte Carlo , b) signal Monte Carlo, and c) data.  Plots are 
logarithmic in $z$.  The box shows the location of the cut used to 
eliminate this background.}
\end{figure}

The most significant source of background comes from $\kmmg$ events in which 
the photon converts to an $e^+\,e^-$ pair in the material upstream of the 
first drift chamber.  Monte Carlo simulations predict almost 50 of these 
events in the signal mass region at this stage of cuts.  Requiring that the 
$e^+\,e^-$ hit separation at the first drift chamber is greater than 2 mm or 
that $M_{ee}>2.75$ MeV/c$^2$ eliminates 99\% of these conversion events, 
while retaining approximately 85\% of the signal (Fig.~\ref{fig:conversions}).

\begin{figure}
\scalebox{0.475}{\includegraphics{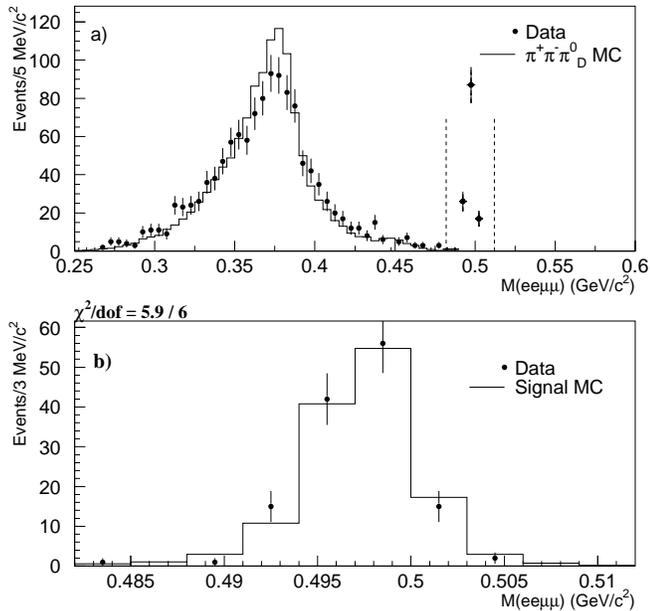}}
\caption{\label{fig:final_mass_overlay} 
a)  $M_{ee\mu\mu}$ for all data (dots) and $\kpmzdal$ decay/punchthrough 
Monte Carlo (histogram) after all cuts.  132 events remain in the signal mass 
region.  The vertical lines indicate the expected signal region. b)
Close--up of the signal mass region.}
\end{figure}

After all cuts, 132 signal events remain (Fig.~\ref{fig:final_mass_overlay}) 
with an estimated background of 0.8 events, 
dominated by $\kmmg$ conversions. The background estimate is
determined from Monte Carlo simulations.
We extract the branching ratio by normalizing to $\kpmzdal$ events, 
which are collected in a similar trigger.  In addition to the 
cuts described, for 
the normalization mode the charged pions are required to strike the 
calorimeter away from the edges of the array (7 cm from the outer 
perimeter, 
5 cm from either beam hole), the energy deposited in the calorimeter by the 
Dalitz photon is required to be greater than 2 GeV, and the mass of the 
reconstructed $\pi^0$ is required to be between 125 and 145 MeV/c$^2$.  
Using these normalization events, and an acceptance of approximately 1.7\% 
determined from Monte Carlo, the total yield of kaon decays within the 
detector from the 1997 and 1999 runs is calculated to be $(6.39\pm 0.04)\times
 10^{11}$, the error being purely statistical.

Because muons are present in the signal mode but not in the normalization 
mode, accurate simulation of the efficiency and threshold of the muon system, 
and a good understanding of multiple scattering through the muon filters, are 
crucial.  These effects are studied using a combination of GEANT simulations 
and calibration muons collected with special magnet and absorber 
configurations.  
The single muon detection efficiency is measured to be over 99\%, determined to within 0.5\% of itself~\cite{gbquinn:thesis}.

Systematic errors in the determination of the number of kaon decays 
and the signal 
acceptance are dominated by the 3.1\% uncertainty in the branching ratio of 
the normalization mode.  Other significant effects include a disagreement 
between data and Monte Carlo in the decay position distribution of 
normalization events from the 1999 run (1.9\%), the rate of accidental 
activity in the detector (1.7\%), and limited Monte Carlo 
statistics (1.1\%).  Other effects include sensitivity to the
vertex quality (1.1\%), the trigger (1.0\%), and the fitting
procedure (0.8\%). The remaining effects result from 
uncertainties in the background and the calibration (0.7\%).
The total systematic error on the branching ratio is 4.6\%.

\begin{figure}
\scalebox{0.475}{\includegraphics{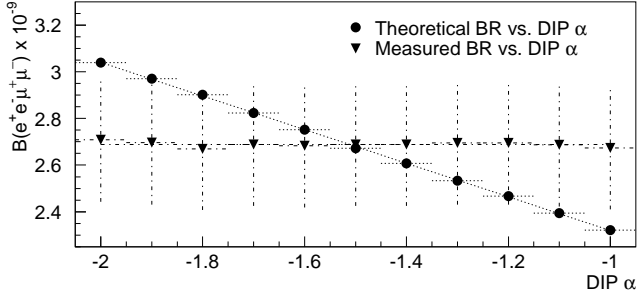}}
\caption{\label{fig:fitalpha_br_tony} Simultaneous determination of
the $\mathcal{B}(\keemm)$ and the DIP $\alpha$ parameter.  The triangles show 
values for the branching ratio, based on 132 signal events with 0.8 
background events, using different values of $\alpha$ to calculate the signal 
acceptance.  The error bars include statistical and systematic errors.  The 
dots show the theoretical dependence of the branching ratio on $\alpha$.  The 
intersection of the two lines provides the results 
$\mathcal{B}(\keemm)=(2.69\pm 0.24_{stat}\pm 0.12_{syst})\times 10^{-9}$ and
$\alpha = -1.51\pm 0.34_{stat}\pm 0.17_{syst}$.  In performing this fit, it 
is assumed that the DIP $\beta$ parameter is 0.}
\end{figure}

Because the branching ratio and $\alpha$ are linked through the form factor, 
the following method is used to determine both of these parameters.  
The measured branching ratio is determined as a function of
$\alpha$, where the $\alpha$ dependence results from the
Monte Carlo acceptance correction. Theoretical 
predictions for the branching ratio are also calculated as a function of 
$\alpha$.  Full single--loop QED radiative corrections are 
included~\cite{tony:pcomm} with a cutoff at $M_{\eemm}^2/M_{K}^2 > 0.95$, and 
the quadratic DIP parameter $\beta$ is assumed to be zero.  
The intersection 
of these two curves, shown in Fig.~\ref{fig:fitalpha_br_tony}, determines 
$\mathcal{B}(\keemm)=(2.69\pm 0.24_{stat}\pm 0.12_{syst})\times 10^{-9}$ 
and $\alpha = -1.51\pm 0.34_{stat}\pm 0.17_{syst}$.  Varying the value of 
$\beta$ over a reasonable range has a negligible effect on the results.

\begin{figure}
\scalebox{0.475}{\includegraphics{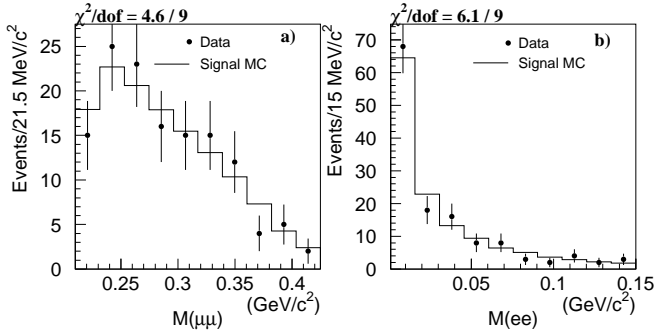}}
\caption{\label{fig:alpha_shape_compare} a) $M_{\mu\mu}$ and b) $M_{ee}$ 
(right) distributions for the 132 signal events, compared to Monte Carlo 
generated with $\alpha=-4.53$.  $\beta$ is zero for both overlays.}
\end{figure}

Because $\alpha$ and $\beta$ are connected to $q_1^2$ and $q_2^2$ 
(Eqn.~\ref{eq:dipff}), independent measurements of the form factor parameters 
can be made by studying the shape of the $M_{\mu\mu}$ and $M_{ee}$ 
distributions of the 132 signal events.  With $\beta$ fixed at 0, the signal 
Monte Carlo is reweighted over a range of values for $\alpha$.  Comparison to 
the data for each value of $\alpha$ leads to a log--likelihood distribution 
that is maximized when $\alpha = -4.53^{+1.81}_{-2.70}$.  A comparison 
between data and Monte Carlo at this value of $\alpha$ is shown in 
Fig.~\ref{fig:alpha_shape_compare}.
Due to the limited statistics, 
we find that this analysis is insensitive to the value of the 
quadratic form factor parameter $\beta$.

A weighted average of the two $\keemm$ measurements of $\alpha$ 
leads to the 
final result $\alpha=-1.59\pm 0.37$, consistent with the two previously 
published values.  Combining the results from $\kmmg$, $\keeee$, and 
$\keemm$ gives a world average of $\alpha = -1.53\pm 0.10$, 
a result that is 
dominated by the measurement from $\kmmg$.

The Bergstr\"{o}m, Mass\'{o}, and Singer (BMS) model for the 
$K_L\gamma^\ast\gamma$ vertex can be generalized to the two--virtual photon 
case~\cite{Bergstrom:1983rj}.  The BMS form factor contains 
only one unknown 
parameter, $\aks$, which can be algebraically related to the DIP parameter 
$\alpha$~\cite{D'Ambrosio:1998jp}.  Using this relation, along with the 
measured value of $\alpha$, we find that $\aks=-0.19\pm 0.11$ from
$K_L\to\ee\mm$. This
is consistent with other KTeV measurements~\cite{Alavi-Harati:2001wd, 
Alavi-Harati:2001ab} as well as those from
the NA48 experiment at CERN~\cite{Fanti:1999}.

\begin{figure}
\scalebox{0.475}{\includegraphics{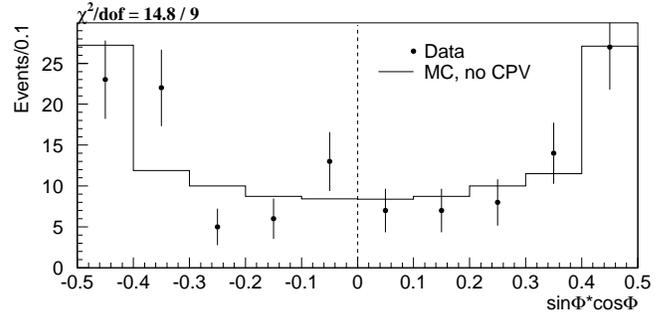}}
\caption{\label{fig:sincos}Angular distribution of the decay products for $\keemm$ events (dots) and Monte Carlo with no $CP-$violation (histogram).}
\end{figure}

Some models of the $\klgg$ vertex allow for $CP-$violating contributions to 
the interaction, the presence of which would lead to an asymmetry in the 
angular distribution of the decay products~\cite{Uy:1991hu}.  The distribution
of events in $\sin\phi\cos\phi$ is shown in Fig.~\ref{fig:sincos}.  $\phi$ is 
the angle between the normals to the electron and muon decay planes in the 
kaon rest frame.
The asymmetry $\mathcal{A}$ is defined by the ratio 
$(N_{+}-N_{-})/(N_{+}+N_{-})$, where $N_{+}\,(N_{-})$ is the acceptance 
corrected number of signal events in the positive (negative) region of 
Fig.~\ref{fig:sincos}.  It is found that $\mathcal{A} = -5.3\pm 12.3\%$, 
which translates to a 90\% confidence limit of $|\mathcal{A}|<25.5\%$.  We 
conclude that no evidence currently exists for a $CP-$violating component 
of the $\klgg$ interaction.

By changing the charge requirement on the lepton pairs, a search can be 
performed for the lepton flavor--violating decay $\kwrongsign$.  After 
imposing a set of analysis cuts otherwise identical to those described 
earlier, no events remain in the signal region.  Four--body phase space 
Monte Carlo was generated to calculate an overall acceptance for this mode 
of approximately 9\%.  This 
leads to a 90\% C.L. limit of $\mathcal{B}(\kwrongsign)<4.12\times 10^{-11}$, 
a factor of three improvement over the previously published 
limit~\cite{Alavi-Harati:2001tk}.

In conclusion, 132 $\keemm$ events have been observed from the 1997 and 1999 
runs of the KTeV experiment, with an estimated background of 0.8 events.  
This signal leads to a measured branching ratio of 
$(2.69\pm 0.24_{stat}\pm 0.12_{syst})\times 10^{-9}$, consistent with 
previously published results.  In the first measurement of the DIP form 
factor parameter $\alpha$ using this decay mode, we find that 
$\alpha = -1.59\pm 0.37$, in agreement with measurements from other modes.  
Because of the limited statistices,
no sensitivity is found to the quadratic DIP parameter $\beta$.  In
order to reduce the error on $\beta$ to be on the order of one, we
estimate that approximately 1,000 times more data would be required.
Furthermore, 
no evidence is found for $CP-$violating contributions to the $\klgg$ 
interaction.  Finally, we constrain the branching ratio of the lepton 
flavor--violating mode $\kwrongsign$ to $<4.12\times 10^{-11}$, at 
90\% confidence.

\bibliography{eemm_prl}

\begin{thebibliography}{15}
\expandafter\ifx\csname natexlab\endcsname\relax\def\natexlab#1{#1}\fi
\expandafter\ifx\csname bibnamefont\endcsname\relax
  \def\bibnamefont#1{#1}\fi
\expandafter\ifx\csname bibfnamefont\endcsname\relax
  \def\bibfnamefont#1{#1}\fi
\expandafter\ifx\csname citenamefont\endcsname\relax
  \def\citenamefont#1{#1}\fi
\expandafter\ifx\csname url\endcsname\relax
  \def\url#1{\texttt{#1}}\fi
\expandafter\ifx\csname urlprefix\endcsname\relax\def\urlprefix{URL }\fi
\providecommand{\bibinfo}[2]{#2}
\providecommand{\eprint}[2][]{\url{#2}}

\bibitem[{\citenamefont{Wolfenstein}(1983)}]{Wolfenstein:1983yz}
\bibinfo{author}{\bibfnamefont{L.}~\bibnamefont{Wolfenstein}},
  \bibinfo{journal}{Phys. Rev. Lett.} \textbf{\bibinfo{volume}{51}},
  \bibinfo{pages}{1945} (\bibinfo{year}{1983}).

\bibitem[{\citenamefont{Belanger and Geng}(1991)}]{Belanger:1991ur}
\bibinfo{author}{\bibfnamefont{G.}~\bibnamefont{Belanger}} \bibnamefont{and}
  \bibinfo{author}{\bibfnamefont{C.~Q.} \bibnamefont{Geng}},
  \bibinfo{journal}{Phys. Rev.} \textbf{\bibinfo{volume}{D43}},
  \bibinfo{pages}{140} (\bibinfo{year}{1991}).

\bibitem[{\citenamefont{Buras and Fleischer}(1998)}]{Buras:1998fb}
\bibinfo{author}{\bibfnamefont{A.~J.} \bibnamefont{Buras}} \bibnamefont{and}
  \bibinfo{author}{\bibfnamefont{R.}~\bibnamefont{Fleischer}},
  \bibinfo{journal}{Adv. Ser. Direct. High Energy Phys.}
  \textbf{\bibinfo{volume}{15}}, \bibinfo{pages}{65} (\bibinfo{year}{1998}),
  \eprint[http://arXiv.org/abs]{hep-ph/9704376}.

\bibitem[{\citenamefont{Uy}(1991)}]{Uy:1991hu}
\bibinfo{author}{\bibfnamefont{Z.~E.~S.} \bibnamefont{Uy}},
  \bibinfo{journal}{Phys. Rev.} \textbf{\bibinfo{volume}{D43}},
  \bibinfo{pages}{802} (\bibinfo{year}{1991}).

\bibitem[{\citenamefont{D'Ambrosio et~al.}(1998)\citenamefont{D'Ambrosio,
  Isidori, and Portol\`{e}s}}]{D'Ambrosio:1998jp}
\bibinfo{author}{\bibfnamefont{G.}~\bibnamefont{D'Ambrosio}},
  \bibinfo{author}{\bibfnamefont{G.}~\bibnamefont{Isidori}}, \bibnamefont{and}
  \bibinfo{author}{\bibfnamefont{J.}~\bibnamefont{Portol\`{e}s}},
  \bibinfo{journal}{Phys. Lett.} \textbf{\bibinfo{volume}{B423}},
  \bibinfo{pages}{385} (\bibinfo{year}{1998}),
  \eprint[http://arXiv.org/abs]{hep-ph/9708326}.

\bibitem[{\citenamefont{Alavi-Harati
  et~al.}(2001{\natexlab{a}})}]{Alavi-Harati:2001wd}
\bibinfo{author}{\bibfnamefont{A.}~\bibnamefont{Alavi-Harati}}
  \bibnamefont{et~al.} (\bibinfo{collaboration}{KTeV}), \bibinfo{journal}{Phys.
  Rev. Lett.} \textbf{\bibinfo{volume}{87}}, \bibinfo{pages}{71801}
  (\bibinfo{year}{2001}{\natexlab{a}}).

\bibitem[{\citenamefont{Alavi-Harati
  et~al.}(2001{\natexlab{b}})}]{Alavi-Harati:2001ab}
\bibinfo{author}{\bibfnamefont{A.}~\bibnamefont{Alavi-Harati}}
  \bibnamefont{et~al.} (\bibinfo{collaboration}{KTeV}), \bibinfo{journal}{Phys.
  Rev. Lett.} \textbf{\bibinfo{volume}{86}}, \bibinfo{pages}{5425}
  (\bibinfo{year}{2001}{\natexlab{b}}),
  \eprint[http://arXiv.org/abs]{hep-ex/0104043}.

\bibitem[{\citenamefont{Alavi-Harati
  et~al.}(2001{\natexlab{c}})}]{Alavi-Harati:2001tk}
\bibinfo{author}{\bibfnamefont{A.}~\bibnamefont{Alavi-Harati}}
  \bibnamefont{et~al.} (\bibinfo{collaboration}{KTeV}), \bibinfo{journal}{Phys.
  Rev. Lett.} \textbf{\bibinfo{volume}{87}}, \bibinfo{pages}{111802}
  (\bibinfo{year}{2001}{\natexlab{c}}),
  \eprint[http://arXiv.org/abs]{hep-ex/0108037}.

\bibitem[{\citenamefont{Alavi-Harati et~al.}(1999)}]{Alavi-Harati:1999zr}
\bibinfo{author}{\bibfnamefont{A.}~\bibnamefont{Alavi-Harati}}
  \bibnamefont{et~al.} (\bibinfo{collaboration}{KTeV}), \bibinfo{journal}{Phys.
  Rev. Lett.} \textbf{\bibinfo{volume}{83}}, \bibinfo{pages}{922}
  (\bibinfo{year}{1999}), \eprint[http://arXiv.org/abs]{hep-ex/9903007}.

\bibitem[{\citenamefont{Alavi-Harati et~al.}(2000)}]{Alavi-Harati:1999hd}
\bibinfo{author}{\bibfnamefont{A.}~\bibnamefont{Alavi-Harati}}
  \bibnamefont{et~al.} (\bibinfo{collaboration}{KTeV}), \bibinfo{journal}{Phys.
  Rev.} \textbf{\bibinfo{volume}{D61}}, \bibinfo{pages}{072006}
  (\bibinfo{year}{2000}), \eprint[http://arXiv.org/abs]{hep-ex/9907014}.

\bibitem[{\citenamefont{Bown et~al.}(1996)}]{Bown:1996qv}
\bibinfo{author}{\bibfnamefont{C.}~\bibnamefont{Bown}} \bibnamefont{et~al.},
  \bibinfo{journal}{Nucl. Instrum. Meth.} \textbf{\bibinfo{volume}{A369}},
  \bibinfo{pages}{248} (\bibinfo{year}{1996}).

\bibitem[{\citenamefont{Quinn}(2000)}]{gbquinn:thesis}
\bibinfo{author}{\bibfnamefont{G.~B.} \bibnamefont{Quinn}}, Ph.D. thesis,
  \bibinfo{school}{The University of Chicago} (\bibinfo{year}{2000}).

\bibitem[{\citenamefont{Barker et~al.}(2002)\citenamefont{Barker, Huang, Toale,
  and Engle}}]{tony:pcomm}
\bibinfo{author}{\bibfnamefont{A.~R.} \bibnamefont{Barker}},
  \bibinfo{author}{\bibfnamefont{H.}~\bibnamefont{Huang}},
  \bibinfo{author}{\bibfnamefont{P.}~\bibnamefont{Toale}}, \bibnamefont{and}
  \bibinfo{author}{\bibfnamefont{J.}~\bibnamefont{Engle}}
  (\bibinfo{year}{2002}), \eprint{hep-ph/0210174}.

\bibitem[{\citenamefont{Bergstr{\"o}m et~al.}(1983)\citenamefont{Bergstr{\"o}m,
  Mass\'{o}, and Singer}}]{Bergstrom:1983rj}
\bibinfo{author}{\bibfnamefont{L.}~\bibnamefont{Bergstr{\"o}m}},
  \bibinfo{author}{\bibfnamefont{E.}~\bibnamefont{Mass\'{o}}},
  \bibnamefont{and} \bibinfo{author}{\bibfnamefont{P.}~\bibnamefont{Singer}},
  \bibinfo{journal}{Phys. Lett.} \textbf{\bibinfo{volume}{B131}},
  \bibinfo{pages}{229} (\bibinfo{year}{1983}).

\bibitem[{\citenamefont{Fanti et~al.}(1999)}]{Fanti:1999}
\bibinfo{author}{\bibfnamefont{V.}~\bibnamefont{Fanti}} \bibnamefont{et~al.}
  (\bibinfo{collaboration}{NA48}), \bibinfo{journal}{Phys. Lett.}
  \textbf{\bibinfo{volume}{B458}}, \bibinfo{pages}{553} (\bibinfo{year}{1999}).

\end{thebibliography}
\end{document}